\documentclass[times,twocolumn,final]{elsarticle}

\usepackage{framed,multirow}

\usepackage{amssymb}
\usepackage{latexsym}

\usepackage{url}
\usepackage{xcolor}

\usepackage{hyperref}

\usepackage{amsmath}
\usepackage{graphicx}
\usepackage{caption}
\usepackage{subcaption}
\usepackage{url}
\usepackage{algorithm}
\usepackage{algpseudocode}
\usepackage{float}
\usepackage{cite}
\usepackage{makecell}
\usepackage{bbm}
\usepackage{comment}

\usepackage{changepage}

\DeclareMathOperator*{\softmax}{softmax}
\DeclareMathOperator*{\argmax}{argmax}
\DeclareMathOperator*{\argmin}{argmin}

\definecolor{newcolor}{rgb}{.8,.349,.1}

\journal{Medical Image Analysis}

\begin{document}


\begin{titlepage}
\begin{center}
      \Large\textbf{Interpretable HER2 scoring by evaluating clinical Guidelines through a weakly supervised, constrained Deep Learning Approach}\\\vspace{0.3cm}
      \large\textit{$^1$Manh Dan Pham, $^2$Cyprien Tilmant, $^3$St\'ephanie Petit, $^4$Isabelle Salmon, \\$^1$Saima Ben Hadj, $^1$Rutger H.J. Fick}\\\vspace{1cm}
      \large\textit{$^1$Tribun Health, 2 Rue du Capitaine Scott, 75015 Paris, France,\\ $^2$GHICL, Lille, France, \\$^3$Xpath Nord, Leulinghem, France,\\ $^4$H\^opital Erasme, Route de Lennik 808 1070, Brussels, Belgium}
\end{center}

\begin{adjustwidth}{50pt}{50pt}
\vspace{1cm}
\section*{Abstract}
The evaluation of the Human Epidermal growth factor Receptor-2 (HER2) expression is an important prognostic biomarker for breast cancer treatment selection. However, HER2 scoring has notoriously high interobserver variability due to stain variations between centers and the need to estimate visually the staining intensity in specific percentages of tumor area.\\
In this paper, focusing on the interpretability of HER2 scoring by a pathologist, we propose a semi-automatic, two-stage deep learning approach that directly evaluates the clinical HER2 guidelines defined by the American Society of Clinical Oncology/ College of American Pathologists (ASCO/CAP). In the first stage, we segment the invasive tumor over the user-indicated Region of Interest (ROI). Then, in the second stage, we classify the tumor tissue into four HER2 classes. For the classification stage, we use weakly supervised, constrained optimization to find a model that classifies cancerous patches such that the tumor surface percentage meets the guidelines specification of each HER2 class. We end the second stage by freezing the model and refining its output logits in a supervised way to all slide labels in the training set. \\
To ensure the quality of our dataset's labels, we conducted a multi-pathologist HER2 scoring consensus. For the assessment of doubtful cases where no consensus was found, our model can help by interpreting its HER2 class percentages output. We achieve a performance of $0.78$ in F1-score on the test set while keeping our model interpretable for the pathologist, hopefully contributing to interpretable AI models in digital pathology.
\end{adjustwidth}
\vspace*{\stretch{1.0}}
\end{titlepage}



\clearpage
\section{Introduction}
Breast cancer is the most prevalent cancer in women worldwide \citep{Sung2021}, accounting for almost one in four cancer cases among women. Between $15\%$ and $20\%$ of breast tumors have higher levels of Human Epidermal growth factor Receptor-2 (HER2) protein. These types of cancer are called HER2-positive and tend to grow and spread faster than HER2-negative cancer, but are sensitive to specific treatments. Thus, correctly assessing the HER2 expression is essential to determine the best treatment for the patient. \\

To stratify patients between HER2-negative and HER2-positive, an immunohistochemistry (IHC) test is performed on a tumor tissue sample. Following the American Society of Clinical Oncology / College of American Pathologists (ASCO/CAP) guidelines \citep{Wolff2018}, a score among 0, $1+$, $2+$, and $3+$ is then attributed by visually inspecting the stain intensities and the surface percentages of differently stained invasive cancer (see Table \ref{guidelines}). However, it has been known for two decades that this visual assessment is prone to a significant inter-observer variability \citep{Thomson2001, Hoang2000}, especially because of stain variations between centers and the need to estimate percentages in the tumor area. \\

With the development of digital pathology, computer-aided solutions have been developed to act as a second opinion for the pathologist \citep{niazi2019digital}. Different studies were conducted to automatically evaluate the HER2 score on immunochemistry (IHC) slides such as \citet{qaiser2019learning} or \citet{chen2021diagnose} and showed a high concordance rate between the artificial intelligence model and the pathologist scoring. However, these methods do not provide a mean for the pathologist to verify or interpret the model's predictions because they do not consider the whole invasive carcinoma but only certain areas to compute the slide's HER2 score. Our work addresses this interpretability issue without compromising our model performance by computing the tumor surface of each HER2 class within the slide and directly implementing the clinical constraints for HER2 scoring in a weakly supervised constrained approach.\\

\begin{figure*}[h]
    \centering
    \includegraphics[width = \linewidth]{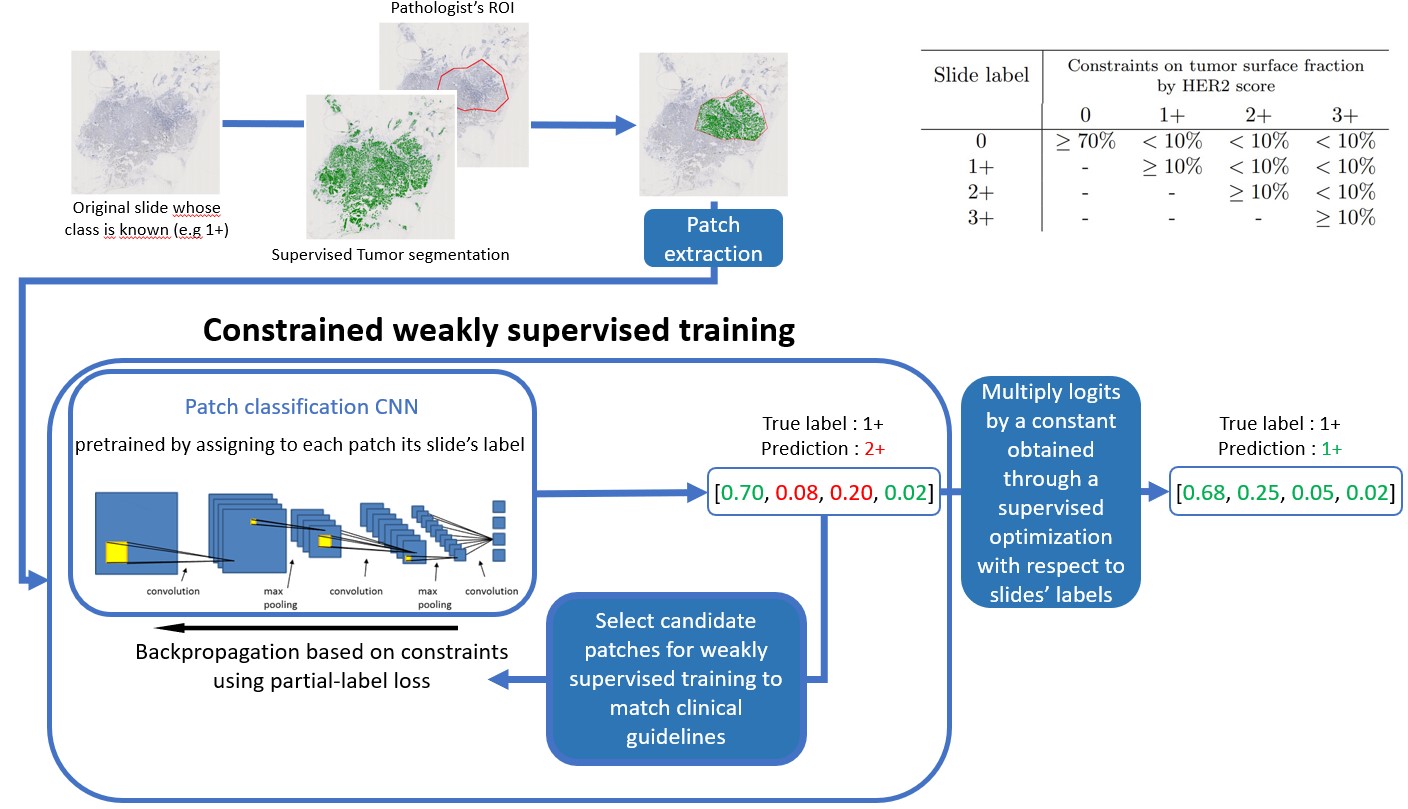}
    \caption{Our end-to-end pipeline for interpretable HER2 slide scoring. The pathologist is asked to choose a Region of Interest (ROI) in which to compute the HER2 score of the slide. Within the ROI, the invasive cancer areas are segmented and patches are extracted within the tumor mask. Then, a model is trained to classify the patches into four different classes corresponding to the HER2 scores. We derive constraints from the 2021 GEFPICS clinical guidelines \citep{Franchet2021} to train the model in a weakly supervised way using only the slides' labels. Once the model is trained, we freeze it and add a supervised optimization step to adjust its logits in a supervised way to the labels of all slides in the training dataset at the same time.}
    \label{fig:graphical_abstract}
\end{figure*}

To allow a pathologist to directly interpret the clinical guidelines in terms of tumor surface percentages, we propose a semi-automatic end-to-end pipeline (see Figure \ref{fig:graphical_abstract}) that provides the tumor surface percentages, together with the spatial class map representation (see Figure \ref{fig:heterogeneous}). The only human intervention is done at the first step, which consists in indicating a Region of Interest (ROI) over which the HER2 expression evaluation will be computed. The purpose of the ROI is to avoid stained tissues that are not taken into account for the evaluation of HER2 expression such as carcinoma in situ or stained benign glands (see Figure \ref{insitu}, \ref{glands}). Within the user-indicated ROI, a segmentation model separates the invasive carcinoma from non-tumor area, and patches around tumor area are extracted. Then, a model is trained to classify the patches into four different \emph{patch} classes (0, $1+$, $2+$, and $3+$) corresponding to locally homogeneous regions for the four HER2 slide scores of the same name in Table \ref{guidelines}. This means we classify the invasive cancer surface as a proxy of the number of invasive cancer cells, on which the guidelines are based. This proxy allows us to avoid segmenting individual nuclei, thus avoiding the need to analyze the slide at high magnification. We derive constraints from the clinical guidelines to train the model in a weakly supervised way so that it classifies cancerous patches to meet the tumor surface percentage constraints of each HER2 class. To further enhance the model's performance once it has been trained, we freeze it and add a model calibration step to adjust its logits in a supervised way to the slides’ labels. To the best of our knowledge, we are the first to use such a weakly supervised approach for directly implementing the clinical constraints for HER2 scoring. \\

In this paper, we first do a literature review on HER2 scoring and the use of weakly supervised learning in histopathology in Section \ref{relatedWorks}. We then detail our constrained weakly supervised approach in Section \ref{method}. The details of the implementation and the results of the experimentation are presented in Sections \ref{implementation} and \ref{results}, followed by a discussion in Section \ref{discussion}.

\begin{table}[htbp]
    \centering
    \caption{Correspondance between ASCO/CAP guidelines and the algorithm constraints. There are some rare heterogeneous staining patterns that are not covered by the ASCO/CAP definitions mentioned in the table. These slides are not considered for training but are discussed in Section \ref{slideAnalysis}.}
    \begin{tabular}{p{0.08\linewidth} p{0.45\linewidth} p{0.35\linewidth}}
        \hline
         HER2 score & 2018 ASCO/CAP guidelines & Algorithm constraints \\
         \hline
         0 & No staining is observed or membrane staining that is incomplete and is faint/barely perceptible in $\leq 10 \%$ of tumor cells
         & $> 70 \%$ of tumor surface is classified as class 0\\
         & & and $<10 \%$ of tumor surface is classified as class 1, 2 or 3 \\
         
         \hline
         1+ & Incomplete membrane staining that is faint/barely perceptible and in $ >10\%$ of tumor cells
         & $\geq 10 \%$ of tumor surface is classified as class 1 \\
         & & and $<10 \%$ of tumor surface is classified as class 2 or 3 \\

         \hline
         2+ & Weak to moderate complete membrane staining observed in  $> 10 \%$ of tumor cells
         &  $ \geq 10 \%$ of tumor surface is classified as class 2 \\
         & & and $< 10 \%$ of tumor surface is classified as class 3 \\
         \hline
         3+ & Circumferential membrane staining that is complete, intense and in $> 10\%$ of tumor cells
         & and $\geq 10 \%$ of tumor surface is classified as class 3 \\
         \hline
    \end{tabular}
    \label{guidelines}
\end{table}


\section{Related works}
\label{relatedWorks}

In HER2 scoring, many methods rely on classifying small patches sampled from the whole slide image (WSI) and then aggregating the patch-level predictions to obtain the slide-level prediction. The patch classification can be done in two ways : fully supervised or weakly supervised, which we describe in the rest of this section. \\

The fully supervised approaches classify all patches within a segmented region obtained by classical image processing techniques. For instance, \citet{vandenberghe2017relevance} extracts all tiles within the slide's foreground, which correspond to the tissue sample. \citet{oliveira2020weakly} uses Otsu's method \citep{otsu1979threshold} to segment cancer tissues from 2+ and 3+ slides, and filters on the HSV value for 0 and 1+ slides, removing patches with the highest H corresponding to background patches. \citet{vandenberghe2017relevance} trained a model to segment and classify individual cancer cells, for which they manually annotated 12 200 cells. \citet{oliveira2020weakly} assigns the slide label to all its patches, which leads to an overclassification bias as we show in Figure \ref{fig:classifMetrics}. To infer the slide label from the patches predictions, \citet{saha2018her2net, vandenberghe2017relevance} and \citet{oliveira2020weakly} directly apply the ASCO/CAP clinical guidelines from their supervised patch classification/segmentation. \\

In the field of weakly supervised methods, the selection of the patches to be evaluated are learned by the model. Inspired from the way pathologists evaluate slides, screening at low magnification followed by a more detailed inspection at high magnification, \citet{qaiser2019learning} propose a deep reinforcement learning approach to automatically identify diagnostically relevant ROI where patches at a magnification of $40\times$ and $20\times$ are extracted. \citet{chen2021diagnose} also implements an automatic multi-scale patch selection by representing the WSI as a tree-structured image and by using an attention module to find discriminative regions. To predict the slide label, they train a shallow classifier from the predicted classes or feature vectors of the patches. Because these weakly supervised approaches do not attribute an HER2 score to all patches with invasive carcinoma, they cannot compute the number of cells of each HER2 score, and hence cannot apply the ASCO/CAP clinical guidelines to infer the slide's HER2 score.  \\

As the cell-wise annotation of HER2 expression requires an extensive amount of annotations from an expert, we prefer to use weak supervision \citep{campanella2019clinical} and let the slide's label induce weakly supervised linear constraints on the patch percentages of each class. In weakly supervised invasive cancer segmentation, \citet{lerousseau2020weakly} have exploited the framework of \citet{campanella2019clinical}. They reframed the top-k patches parameter for assigning pseudo-labels to patches as two control parameters which indicate the percentage of tumor and normal tissue that should be present in the slide according to the pathology report. Dictating specific constraints on tumor surface percentages of different HER2 classes can be seen as a multi-class implementation of the weakly supervised segmentation problem, where we assign pseudo-labels to the top-K \% patches with probabilities of certain classes, but only for classes that break linear constraints dictated by the clinical guidelines. \\
Thus, our approach leverages the advantages of both fully and weakly supervised HER2 scoring paradigms: our weakly supervised training does not require extensive expert annotations and still provides an interpretable output for the slide's label as we directly use the ASCO/CAP clinical guidelines for training our model and predicting the slide's label.

\section{Methods}
\label{method}
\subsection{Dataset}
Our proposed framework is performed on HER2 IHC stained slides of breast tissue. The dataset is composed of 370 (WSI) coming from different sources, including 270 WSI from two different scanners of Erasme Hospital (Hamamatsu NanoZoomer S360 and Hamamatsu HT-C9600-12), 50 WSI from Warwick HER2 scoring contest training set \citep{qaiser2018her} scanned with the Hamamatsu NanoZoomer C9600, and 50 WSI from the Academia and Industry Collaboration for Digital Pathology (AIDPATH) database. Erasme's and Warwick's slides were provided with an IHC score (0, 1+, 2+ or 3+), Erasme scoring being conducted using the 2018 ASCO/CAP guidelines, such that $0$ and $1+$  slides were both considered HER2-negative. Slides from AIDPATH database only have the clinical outcome that is HER2 negative, positive and equivocal with respectively 37, 7, and 6 slides. The numbers of slides per class and scanner for Erasme and Warwick datasets are summarized in table \ref{tab:database}.

\subsection{Invasive Carcinoma Segmentation Annotations}
To train the tumor segmentation model, we asked a pathologist to annotate tissue area on 71 WSI from Erasme and AIDPATH using the Calopix software\footnote{\url{https://www.tribun.health/calopix}}. On Erasme dataset, we annotated 20 mm² of class 0 from 15 slides, 23 mm² of class 1+ from 16 slides, 16 mm² of class 2+ from 6 slides, and 84 mm² of class 3+ from 6 slides. On the AIDPATH dataset, 21 mm² were annotated from 22 HER2-negative slides, 9 mm² from 3 equivocal slides, and 5 mm² from 3 HER2-positive slides. The annotations were done in incremental steps according to the model performance for each class until a target performance of around 0.9 in F1-score was reached across all classes.

\begin{table}[htbp]
    \centering
    \caption{Number of slides of each class from each dataset. Note that the Erasme datasets are private and we use the training set of the Warwick HER2 challenge as the test set.}
    \begin{tabular}{l|l| cccc}
         \hline
         \multicolumn{1}{l}{Dataset} &  \multicolumn{1}{|l}{Scanner} & \multicolumn{4}{|c}{\thead{Number of slides \\ per HER2 score}} \\

         & & $0$ & $1+$ & $2+$ & $3+$ \\
         \hline
         Erasme & Hamamatsu C9600 & 23 & 36 & 63 & 30  \\
         Erasme & Hamamatsu S360  & 26 & 40 & 46 & 6 \\
         Warwick & Hamamatsu C9600 & 13 & 12 & 12 & 13 \\
    \end{tabular}

    \label{tab:database}
\end{table}

\begin{figure}[h]
    \centering
    \includegraphics[width=\linewidth]{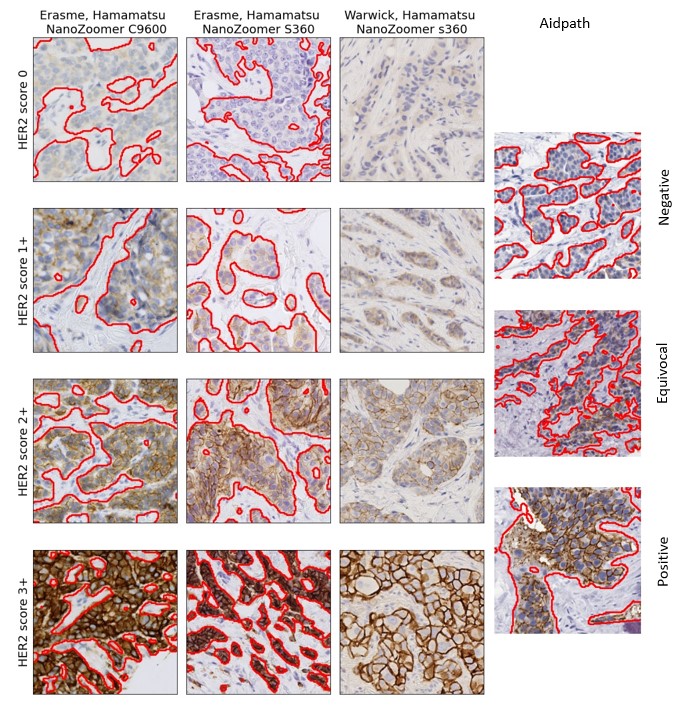}
    \caption{Patches with different HER2 expressions for all data sources (magnification $10 \times$, $1 \mu m /pixel$). The tumoral areas, shown in red, were annotated by a pathologist. Note that the Warwick dataset was not annotated as it was not used for training the tumor segmentation model and that AIDPATH dataset does not have precise IHC scores but only the clinical outcome (HER2 negative, equivocal, or positive).}
    \label{fig:patches}
\end{figure}

\begin{figure}[h]
    \centering
    \begin{subfigure}{0.3\linewidth}
    \centering
        \includegraphics[width=\linewidth]{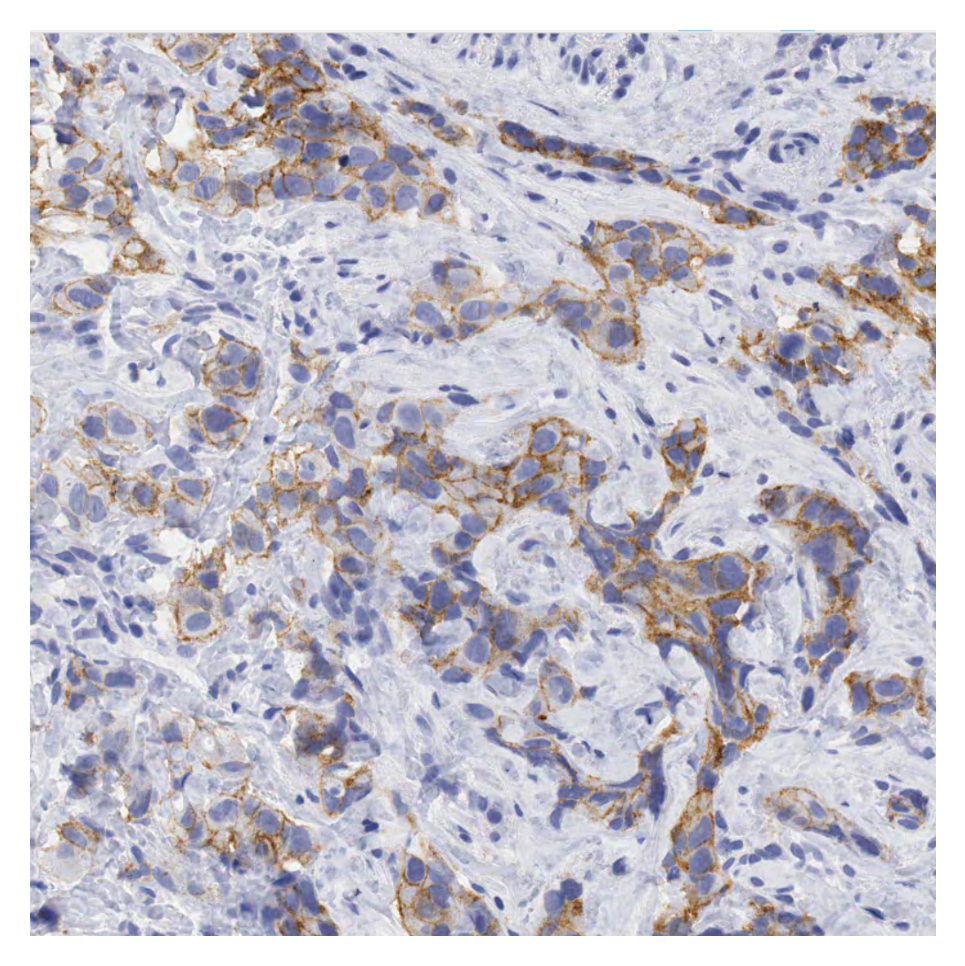}
        \caption{Invasive carcinoma}
    \end{subfigure}
    \begin{subfigure}{0.3\linewidth}
        \centering
        \includegraphics[width=\linewidth]{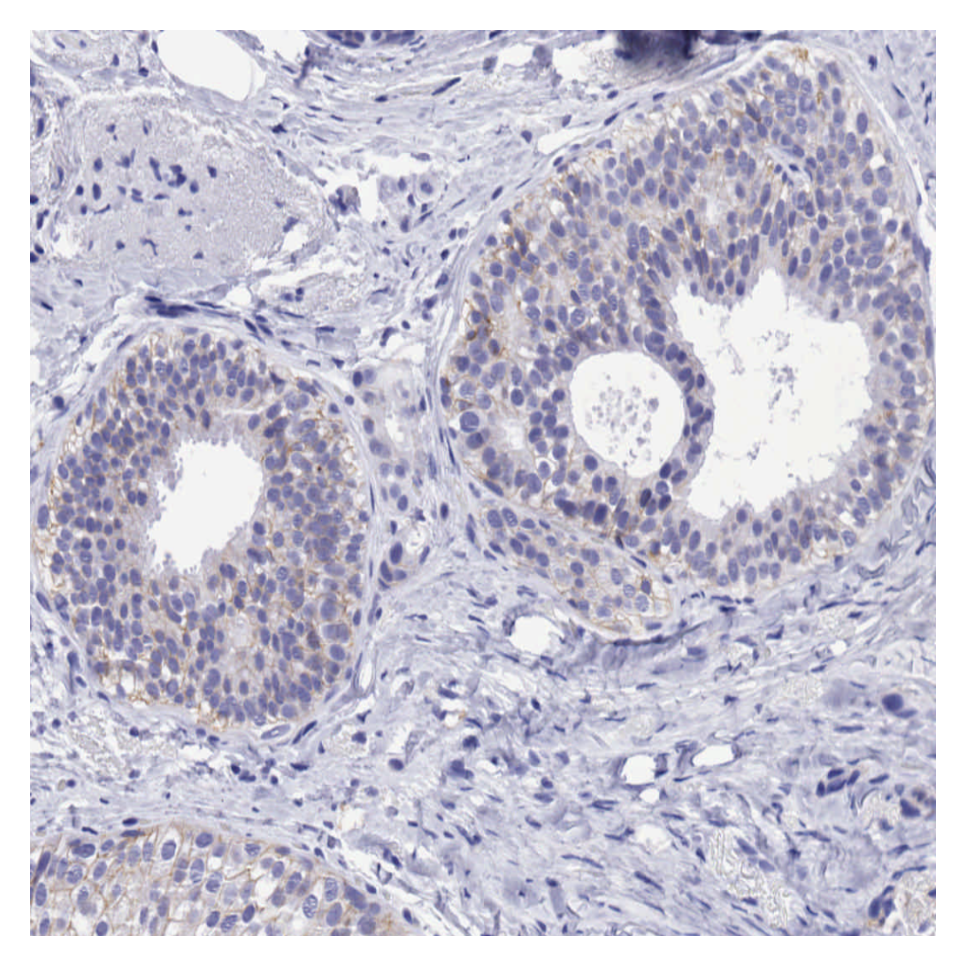}
        \caption{Carcinoma in situ}
        \label{insitu}
    \end{subfigure}
    \begin{subfigure}{0.3\linewidth}
        \centering
        \includegraphics[width=\linewidth]{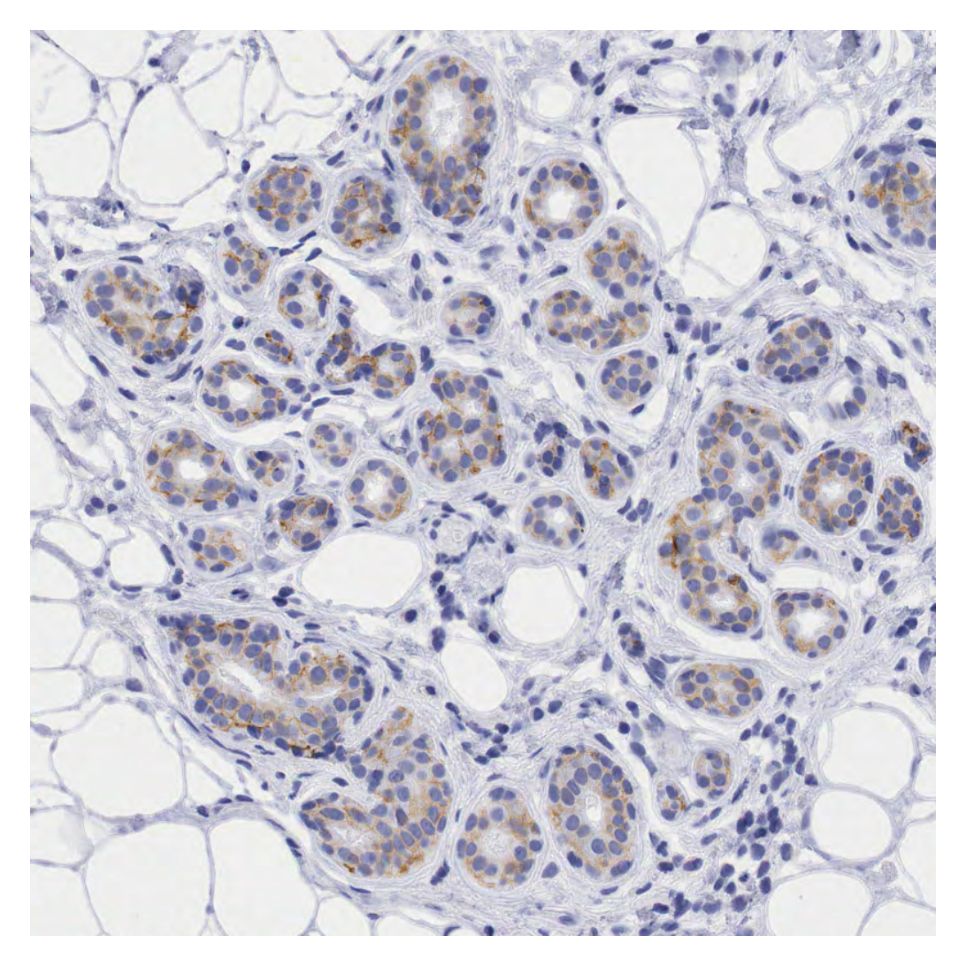}
        \caption{Stained benign glands}
        \label{glands}
    \end{subfigure}
    \caption{Different types of stained tissues. Only invasive carcinoma must be taken into account for the evaluation of the slide's HER2 score. The pipeline is launched only within a Region Of Interest (ROI) selected by the pathologist which excludes undesirable structures such as carcinoma in situ, stained benign cells or artifacts.}
    \label{fig:centers}
\end{figure}

\subsection{Labeling HER2 slides using multi-pathologist consensus for GEFPICS 2021 guidelines}\label{subsec:consensus_scoring}
The evaluation of HER2 expression is subject to a significant inter-observer variability \citep{Thomson2001} because staining intensities vary from one center to another (see e.g. figure \ref{fig:centers}). Moreover, the guidelines for evaluating the HER2 expression requires to assess specific percentages of invasive carcinoma cells across the whole slide, which can only be eyeballed in practice. This slide scoring variability not only has a detrimental impact on patient treatment, but also can be seen as label noise, prohibiting us from accurately modeling the relationship between the HER2 stain intensity and the slide's HER2 score. \\

\begin{figure}[h]
    \centering
    \includegraphics[width=\linewidth]{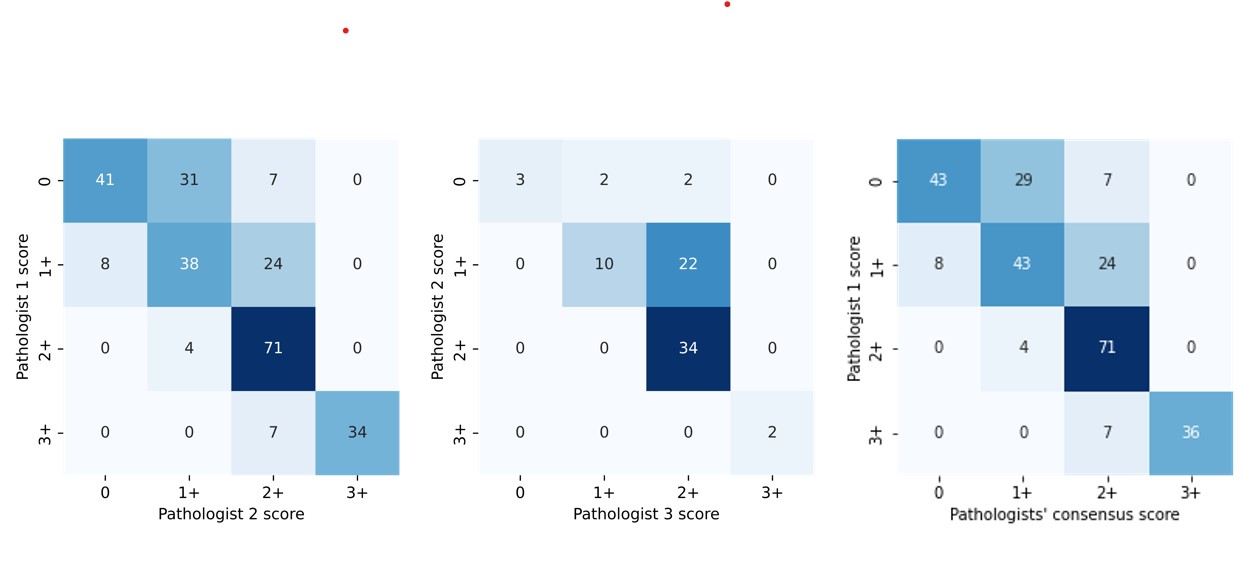}
    \caption{Confusion matrices on the slide HER2 score between the different pathologists. From left to right: confusion matrix between pathologists 1 and 2 who scored all slides independently, confusion matrix between pathologists 2 and 3 for discordant cases between pathologists 1 and 2, confusion matrix between pathologist 1 and the consensus of pathologists 2 and 3.}
    \label{fig:pat1}
\end{figure}

The recent development of drugs targeting HER2-low cancer \citep{modi2020antitumor} has pushed clinical guidelines to refine the practices of HER2 assessment between HER2-negative and HER2-low cases. For instance, in France, 2021 GEFPICS clinical guidelines \citep{Franchet2021} now include different clinical decisions for HER-low cases which are $1+$ and $2+$ FISH-negative cases, HER2-negative only corresponding to $0$ cases. Some studies, such as \citet{moutafi2022quantitative}, suggest new staining techniques to better stratify the lower ranges of HER2 expressions, which highlights the difficulty of differentiating 0 from $1+$ cases with the current conventional assays. To reduce label noise due to inter-observer variability, we initiated a multi-pathologist labeling of Erasme's dataset keeping the 2021 GEFPICS recommendations in mind. First, two pathologists scored independently all Erasme's slides. Then, for the cases where the pathologists' labels differed, we had a third pathologist to independently score the disagreement cases. For cases where the scoring was indicated as uncertain by all pathologists, or where all three pathologists had a different labels, we had a consensus meeting to find the final label. We did not annotate Warwick and AIDPATH datasets as we did not have the control slides to guide the scoring. For some boundary cases or heterogeneous cases where the pathologists could not decide with confidence on the final label, we set these slides aside to be evaluated later with our trained model, and discuss them in the Section \ref{slideAnalysis}. \\
Figure \ref{fig:pat1} shows the differences in scoring between the pathologists. The first confusion matrix on the left displays the scoring of pathologists 1 and 2 that were made independently. The review of the discordant cases by pathologist 3 is compared to the annotations of pathologist 2 in the central confusion matrix. The matrix on the right compares pathologist 1 labels to the labels assigned after the consensus meeting between pathologists 2 and 3.

\subsection{invasive carcinoma segmentation}
To separate benign tissue from cancerous tissue, we train a model to segment all tumor pixels from the WSI on the annotated slides from Erasme and AIDPATH datasets. Patches of size $256 \times 256$ are extracted at a magnification of $10\times$ ($1 \mu m$ / pixel) from the annotated areas. The patches are randomly split into 80\%, 10\%, 10\% for training, validation and test set. We perform a cross-validation with 3 different splits to evaluate the model. The splits are strictly done at the slide level, meaning that all patches from the same slide end in the same set. For the training, we apply different augmentations to the patches using the library albumentations \citep{albumentations} including rotations, flipping, brightness and contrast variation, blurring and, hue and saturation shift. \\
We use a U-net \citep{ronneberger2015u} of depth 4 with a Densenet121 \citep{huang2017densely} as the encoder pretrained on ImageNet \citep{deng2009imagenet}. The U-net is fine-tuned using stochastic gradient descent (SGD) with a Nesterov momentum of $0.9$ as the optimizer with an initial learning rate of $10^{-4}$. We use a weighted focal loss to account for class imbalance between tumor pixels and non-tumor pixels. We also address the imbalance of domains and slides' HER2 scores by sampling equally the patches with respect to the domains and HER2 score at each epoch. The model is trained for 100 epochs with an early stopping based on the validation loss.

\subsection{Multi-stage HER2 patch classification}
In this section, we first introduce the partial-label loss (section \ref{partialloss}) used for the weakly supervised training of the model. We then detail the 3 steps of our optimization process which are the pretraining of the neural network in Section \ref{pretrain}, the weakly supervised training in Section \ref{ws}, then the fully supervised optimization in Section \ref{finalopt}.
The model's inputs are patches extracted at size $64\times64$ at magnification $10\times$ ($1 \mu m$ / pixel). Only patches with more than $10 \%$ of invasive carcinoma surface are kept for training and evaluation, which results in $96 \%$ on average of the total surface of invasive carcinoma within the user-indicated ROI being treated.

\subsubsection{Defining Partial-Label Cross-Entropy Loss \label{partialloss}}

In the Multiple Instance Learning (MIL) paradigm, only the bag's label (the slide) is known while its instances' labels (the patches) remain unknown during training. However, the bag's label can give information on its instances' label. In the case of HER2 scoring, the ASCO/CAP clinical guidelines provide such information (see Table \ref{guidelines}) about the proportions of stained invasive carcinoma cells of each class. We take the tumor surface within classified patches as a proxy for actually having the number of cells. A HER2 slide can be misclassified in two ways: it can either be \emph{over}classified because there are too many patches classified as a class higher than the known class, or it can be \emph{under}classified because there are not enough patches classified as the known class. For instance, for a slide of known class 1+, if the model predicts the tumor surface fractions as $V = [ 0.50, 0.36, 0.14, 0]$ for class $0$, $1+$, $2+$, $3+$, we must reduce by at least 4\% the amount of tumor surface of class 2+ to be in agreement with the slide's label. We know that the $4\%$ in excess of $2+$ patches cannot be of class $2+$ but their true label can be any of the other classes. Thus, we are in the case of partial labeling where for some patch $x$, its true label $y$ is unknown but a set of admissible labels $G$ is known such that $y \in G$. \citet{PartialLoss} introduce a partial-label cross-entropy (CE) loss to learn from these partial labels which is defined as follows. \\
Let $\hat{y}$ be the model prediction after a softmax layer, $G$ the set of admissible class, the partial-label CE loss is defined as :
\begin{equation}
    \mathcal{L}_{part}(\hat{y}) = \mathcal{L}_{CE}(\bar{y}, \hat{y}) = \sum_{j=1}^L - \bar{y}_j log(\hat{y}_j)
\end{equation}

where $\bar{y}$ is the pseudo-label whose expression is:
\begin{equation}
    \forall j \in \llbracket 1, L \rrbracket, \bar{y}_j = 
    \left\{
    \begin{array}{ll}
        \hat{y}_j + \dfrac{1}{|G|} \sum_{k \notin G} \hat{y}_k & \text{ if }j \in G   \\
        0 & \text{ if } j \notin G 
    \end{array}
    \right.
\end{equation}
A key property of the partial-label CE loss is its gradient neutrality towards the admissible classes, meaning that the patch is pushed \textit{equally} towards each admissible class, as we have no information to prioritize one over the others.

\subsubsection{Baseline model used as pretraining \label{pretrain}}
To pretrain our classification model to extract features that are relevant for the HER2 IHC domain, we train a network in a supervised way by assigning to every patch its slide label as done by \citet{oliveira2020weakly}. This approach introduces an overclassification bias because it assumes that the tumor in the entire slide is homogeneous and only consists of the slide's label. In practice, classes lower than the slide's class can exist in significant amounts in the same slide as shown in Figure \ref{fig:kde}, as long as the proportions of the ASCO/CAP clinical guidelines are respected. So this supervised way of training is an approximation of the patches' true labels. Although the given labels to the patches do not necessarily match their true labels, this approach still allows the network to learn appropriate feature extraction based on histopathological images. We trained a Resnet18 pretrained on ImageNet  using SGD with Nesterov momentum of $0.9$ with an initial learning rate of $10^{-3}$ for 100 epochs with a batch size of $512$. The training was stopped if the validation accuracy did not improve during 20 epochs.

\subsubsection{Constrained weakly supervised classification \label{ws}}
Inspired by the ASCO clinical guidelines, we adopt a weakly supervised approach for classifying tumor patches constrained on the proportions of tumor of each class with respect to the slide's label. The training pipeline for the patch classification model is as follows (see Figure \ref{fig:graphical_abstract}): \\
At each epoch, we first do an inference loop over all slides in the training set where all the patches within the ROI and invasive carcinoma segmentation mask are classified. \\
Then, the percentage of the stained tumor surface of each class is computed for every slide. Wrong predictions at the slide level mean that the patch classification model does not respect the clinical guidelines. Our goal is to encourage the classification model to classify patches such that the percentage-based constraints at the slide level are satisfied. Thus, for each epoch, we must select patches from classes that were over- and under-represented for training slides of all classes, and push them away or toward the class in question. We base this selection procedure on the predicted class probability of each class like done by \citet{campanella2019clinical} in their weakly supervised approach:  we push away patches whose probability for the over-represented class is lowest, and inversely push patches with the highest probability of the under-represented (but are not classified as that class) towards that class. This patch selection process defines the training set for the epoch. Thus, the training set for each epoch changes depending on which constraints on which slides are broken. \\

More formally, we note $f_{\theta}$ the network for patch classification where $\theta$ are the parameters of the model. Let consider a slide $X = \{x_i\}_{i=0}^{N-1}$ of class $Y \in \llbracket 0, 3 \rrbracket$, with $N$ its number of patches. For all $i \in \llbracket 0, N-1 \rrbracket$, we note:
\begin{itemize}
    \item $v_i \in [0,1]$ the proportion of tumor pixels in the patch $x_i$, normalized with respect to the total tumor surface in the slide's ROI annotation, such that $\sum_{0}^{N-1} v_i = 1$.
    \item $\hat{y}_i = \argmax \left( \softmax (f_{\theta}(x_i)) \right) \in \llbracket 0, 3 \rrbracket$ the predicted class of the patch $x_i$.
\end{itemize}
Let us define the upper and lower thresholds matrices based on the ASCO clinical guidelines (see Table \ref{guidelines}). 
\begin{equation}
    L = 
\begin{pmatrix}
0.7 & 0 & 0 & 0 \\
0 & 0.1 & 0 & 0 \\
0 & 0 & 0.1 & 0 \\
0 & 0 & 0 & 0.1
\end{pmatrix}
\qquad
U = 
\begin{pmatrix}
0 & 0.1 & 0.1 & 0.1 \\
0 & 0 & 0.1 & 0.1 \\
0 & 0 & 0 & 0.1 \\
0 & 0 & 0 & 0
\end{pmatrix}
\label{constraint_matrices}
\end{equation}
The matrix L indicates class fractions that a slide must have \emph{at least} to be of a certain class and $U$ the fractions to which each tumor surface of a certain class must be lower, otherwise, the slide would be overclassified. \\

In this section, we use the following notation for conditional sum. Given a set $E$ and some scalar $(\alpha_i)_{1 \leq i \leq N}$,  $ \sum\limits_{\substack{ 0 \leq i \leq N \\ \text{if }i \in E}} \alpha_i$ means that we sum only the $\alpha_i$ whose index $i$ is in the set $E$. \\

We introduce the class fractions vector $V = (V_c)_{0 \leq c \leq 3}$ defined by : 
\begin{equation}
    \forall c \in \llbracket 0, 3 \rrbracket, V_c = \sum_{\substack{0 \leq i \leq N-1  \\ \textrm{if } \hat{y}_i = c}} v_i 
\end{equation}
such that $0 \leq V_c \leq 1$ and $\sum_{c=0}^3 V_c = 1$. \\
The constraints are written as follows: \\
\begin{equation}
\label{upper constraints}
    \forall c \in \llbracket 0,3 \rrbracket, V_c < U_{Y, c} 
\end{equation}

\begin{equation}
\label{lower constraints}
    V_Y \geq L_{Y, Y}
\end{equation}

If one of the upper constraints is broken, it means that there exists a class higher than the slide's known class, ie. $c > Y$, such that $V_c - U_{Y, c} \geq 0$ which is the surface proportion of the invasive carcinoma in excess of class $c$. Thus, we must change the prediction of $V_c - U_{Y, c}$ of the tumor surface proportion to meet the clinical guidelines. For this, we sort the patches classified as class $c$ by the model probability for that class : 
\begin{equation}
0 \leq f_{\theta}(x_{i_1})_c \leq \dotsc \leq f_{\theta}(x_{i_N})_c \leq 1
\end{equation}
For a set of classes $E$, we define the cumulative distribution function (CDF) of invasive carcinoma surface proportion of patches classified in $E$ by :
\begin{equation}
    \begin{array}{lrl}
     CDF_E \colon & \llbracket 0, N-1 \rrbracket & \to [0, V_c] \\
     & n &\mapsto \sum\limits_{\substack{0 \leq i \leq n  \\ \textrm{if } \hat{y}_{i_k} \in E}} v_{i_k} 
    \end{array}
\end{equation}

The patches with the \emph{lowest} probabilities are selected until the tumor surface in the remaining patches is less than the exceeded constraint. We choose the lowest probability ones because they are the ones that are most likely to be misclassified. Let $n^u_c$ be the upper cutoff index for class $c$, meaning that patches $x_{i_k}$ for $k \leq n^u_c$ are going to be added to the training set so they are pushed away from class $c$. We take : 

\begin{equation}
\begin{aligned}
n^u_c = & \argmin_n \quad  CDF_c(n) \\
& \textrm{s.t.} \quad CDF_c(n) \geq V_c - U_{Y,c}\\
\end{aligned}
\end{equation}
Thus, the patches selected to be added to the training set because the upper constraints are broken are as follows :
\begin{equation}
S^u_c = \left\{ x_{i_n} \mid n \in \llbracket 0, n^u_j \rrbracket \right\} 
\vspace{0.5cm}
\end{equation}

If the lower constraint is not respected, it means that $L_{Y,Y} - V_Y > 0$ which is the missing tumor surface proportion of the slide's class. Patches from neighboring classes are selected until the tumor surface of the slide's known class is higher than the lower constraints. The patches' selection is based on their probability to belong to the slide's class, the ones with \emph{highest} probabilities being selected:
\begin{equation}
0 \leq f_{\theta}(x_{i_1})_Y \leq \dotsc \leq f_{\theta}(x_{i_N})_Y \leq 1
\end{equation}
Let $n_l$ be the lower cutoff index, meaning that patches $x_{i_k}$ for $k \geq n_l$ are going to be added to the training set so they are pushed toward the slide's known class $Y$. 
\begin{equation}
\begin{aligned}
n^l = & \argmin_n \quad  1 - CDF_{\{Y+1, Y-1\}}(n) \\
& \textrm{s.t.} \quad 1 - CDF_{\{Y+1, Y-1\}}(n) \geq L_{Y,Y} - P_Y\\
\end{aligned}
\end{equation}
Thus the patches selected to be added to the training set because the lower constraints are broken are :
\begin{equation}
S^l = \left\{x_{i_n} \mid  n \in \llbracket n^l, N \rrbracket \right\} 
\end{equation}
Taken together the selected patches for all broken upper and lower constraints, the subset of patches from the slide used as the training set is defined by:
\begin{equation}
S = \bigcup_{c=0}^3 S^u_c \cup S^l
\end{equation}

Our approach is motivated by \citet{campanella2019clinical} who select patches with the highest probability and attach a pseudo-label to it based on the slide label. However, we apply their approach class-wise and do the opposite (selecting patches with the lowest probability) for broken upper constraints. \\
Let $S^u = \bigcup_{c=0}^3 S^u_c$ be the set of patches coming from upper constraints and $S_l$ the set of patches coming from lower constraints such that $S = S^u \cup S^l$. To enforce the model prediction to respect the ASCO guidelines, we want to change the model prediction for patches in $S^u$. We want to assign pseudo-labels to the selected patches such that their probabilities move away from their current exceeding classes but we do not know to which class they belong. Thus, the partial-label CE loss is applied to push equally these patches to their neighboring classes. \\
For patches coming from $S^l$, the goal being to push them towards the slide's class $Y$ to meet the lower-bound condition, a classical cross-entropy is applied with respect to the slide's label $Y$. With $N_S$ being the number of slides, the optimization problem for the epoch is written : 
\begin{equation}
    \argmin\limits_\theta  \sum_{i=1}^{N_S} \left( \sum_{x \in S^u(i)} \mathcal{L}_{CE}(Y_i, f_\theta(x)) + \sum_{x \in S^l(i)} \mathcal{L}_{part}(f_\theta(x)) \right)
\end{equation}
where $S^u(i)$ are the patches that are overclassified, and $S^l(i)$ the patches missing from the slide's class for slide $i$.

\subsubsection{Supervised optimization on the slides' labels \label{finalopt}}

To further increase the performance of our end-to-end pipeline, we perform one more optimization on the neural network logits output. After the weakly supervised optimization, the weights of the classification model are frozen and its logits are finetuned in a supervised way to the labels of all the slides in the training set. For misclassified slides, we aim at minimizing the distance between the thresholds and the proportions of tumor of the classes in excess or missing.\\
Let $N_S$ be the number of slides in the training dataset, $n_i$ the number of patches for slide $i$, and $N = \sum_{i=1}^{N_S} n_i$ the total number of patches for all slides in the training dataset. Let $M \in \mathbb{R}^{N \times 4}$ be the matrix obtained by vertically stacking the logits vectors output by the network for all patches. Although the number of patches $N$ is a very large number, the matrix $M$ fits as once in the memory because each patch is now compressed to its four logits.
\begin{equation}
M = 
\begin{pmatrix}
    L_{1,0} & \dotsm & L_{1,3} \\
    \vdots &    & \vdots \\
    L_{n_1,0} & \dotsm & L_{n_1,3} \\
    \vdots &    & \vdots \\
    L_{N,0} & \dotsm & L_{N,3}
\end{pmatrix}
\in \mathbb{R}^{N \times 4}
\end{equation}
Let us define $\boldsymbol{\alpha} = (\alpha_0, \alpha_1, \alpha_2, \alpha_3) \in \mathbb{R}^4$ the parameters to optimize. They are initialized at the value $(1, 1, 1, 1)$ meaning that the model output logits are not initially modified. The patches' HER2 classes are predicted by the formula:
\begin{equation}
\boldsymbol{\hat{y}}
=
\begin{pmatrix}
\hat{y}_1 \\
\vdots \\
\hat{y}_N
\end{pmatrix}
=
\argmax \biggl(\softmax \Bigl( M*Diag(\alpha_0, \alpha_1, \alpha_2, \alpha_3) \Bigr) \biggr) 
\end{equation}
From the patches' HER2 score predictions $\boldsymbol{\hat{y}}$ and the number of tumor pixels per patch, we can compute the proportion of tumor surface for each slide. For a given slide $i \in \llbracket 1,N_s \rrbracket$, let us note $(v_i)_{1 \leq i \leq n_i}$ the proportion of tumor pixels in its patches given by the invasive carcinoma segmentation model. The proportion of tumor surface $V_{i,c}$ of class $c \in \llbracket 0, 3 \rrbracket$ in the slide is given by:
\begin{equation}
    V_{i,c} = \dfrac{\sum_{k=1}^{n_i} v_k \mathbbm{1}[\hat{y}_k = c]}{\sum_{k=1}^{n_i} v_k}
\end{equation}
Note that for any slide $i \in \llbracket 1,N_s \rrbracket$, $\sum_{c=0}^3 V_{i,c} = 1$. \\
Let $V$ be the matrix of the proportion of tumor surface of all slides in the training set:
\begin{equation}
V = 
\begin{pmatrix}
V_{1,0} & \dotsm & V_{1,3} \\
\vdots & & \vdots \\
V_{N_S,0} & \dotsm & V_{N_S,3}
\end{pmatrix}
\in [0, 1] ^{N_S \times 4}
\end{equation}
Using the upper and lower thresholds matrices based on the ASCO clinical guidelines defined in Eq. \ref{constraint_matrices}, the supervised optimization on the slide's labels to minimize the distance between the thresholds and the tumor surface proportion is :
\begin{equation}
\argmin\limits_{\alpha} \sum_{i=1}^{N_s}  \left[ \left(L_{\hat{y}_i, \hat{y}_i} - V_{i,\hat{y}_i}  \right)^+ + \sum_{\substack{0 \leq c \leq 3 \\ \textrm{if } c > \hat{y}_i}} \left(V_{i,c} - U_{\hat{y}_i, c} \right)^+   \right]
\end{equation}

where $x^+ = \max(0, x)$ denotes the positive part of $x$.

\section{Implementation}
\label{implementation}
The experiments were done using PyTorch $1.7.1$ \citep{paszke2019pytorch} on an HP Z2 G4 Tower Workstation equipped with an NVIDIA GeForce RTX 2070 GPU and an Intel Core i7-8700 CPU. For the fully supervised optimization with respect to the slides' labels, we used the Scipy $1.6.2$ \citep{virtanen2020scipy} implementation of Broyden–Fletcher–Goldfarb–Shanno (BFGS) algorithm to minimize our cost function.

\section{Results}
\label{results}

In this section, we show qualitative and quantitative results of the invasive carcinoma segmentation model and the HER2 patch classification model. For the latter, we compare the performances of the model at each stage (pretrained, weakly supervised, and finetuned) to evaluate the performance improvements made by each of them. 

\subsection{Invasive carcinoma segmentation}
\begin{figure}[h]
    \centering
    \includegraphics[width=0.7\linewidth]{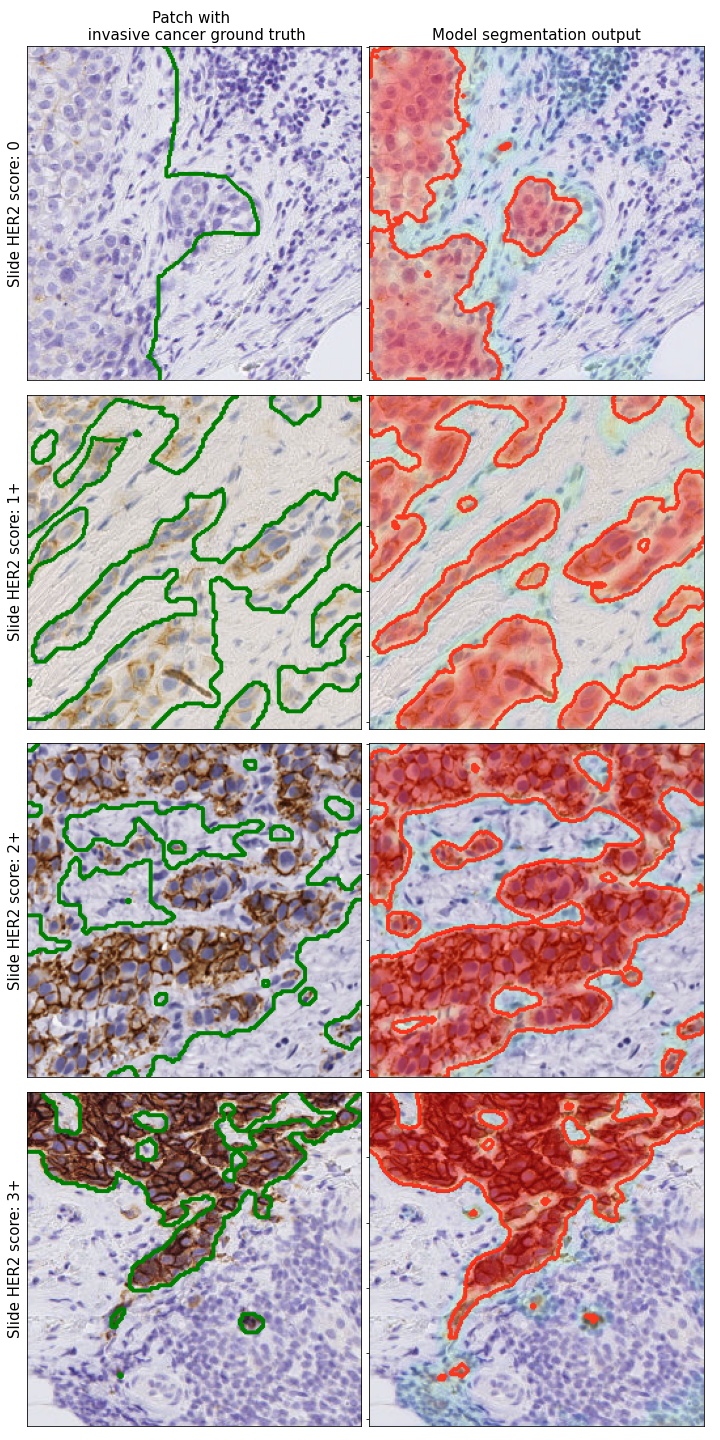}
    \caption{Heatmap of the binary segmentation network output in the right column. The ground truth obtained from the pathologist's annotated is shown in green in the left column. The images are shown in magnification $10 \times$ ($1 \mu m /pixel$). The first row shows a case where the model output is more precise than the manual annotations.}
    \label{fig:seg_heatmap}
\end{figure}

The pixel-wise tumor segmentation network is evaluated on patches split by their slides' HER2 score to account for the different staining intensities. The qualitative results for each HER2 score is shown in Figure \ref{fig:seg_heatmap}. To assess quantitatively the model performance, we use the Dice score, precision, and recall that are computed separately for each HER2 score in table \ref{tab:segmentation_results}. For a given class (tumor / non-tumor), let us note $TP$, $FP$, and $FN$ the numbers of true positive, false positive and false negative pixels. The above metrics are defined by:
\begin{equation}
    \begin{array}{c}
        Precision = \dfrac{TP}{TP + FP}  \\ [10pt]
        Recall = \dfrac{TP}{TP + FN}   \\
        Dice = \dfrac{2}{Precision^{-1} + Recall^{-1}} = \dfrac{2 TP}{2TP + FP + FN}
    \end{array}
\end{equation}


Our segmentation network achieves a performance of above $0.82$ in Dice score on the test set on all classes, 0 and 1+ being the most challenging as the intensity of the staining is weaker.

\begin{table}[h]
    \caption{Pixel-wise metrics of the binary tumor segmentation model on the training, validation and test sets stratified by the slide HER2 score.}
    \begin{subcaption}{Training set}
        \begin{tabular}{c|c c c}
         \hline
         \makecell{Slide \\ HER2 score} & Precision & Recall & Dice score \\
         \hline
         $0$ &  $.886 \pm .029$ & $.861 \pm .052$ & $.872 \pm .014$\\
         $1+$ &  $.853 \pm .039$ & $.925 \pm .025$ & $.887 \pm .016$\\
         $2+$ &  $.815 \pm .028$ & $.875 \pm .025$ & $.844 \pm .006$\\
         $3+$ &  $.901 \pm .019$ & $.950 \pm .017$ & $.925 \pm .005$\\
         \hline
    \end{tabular}
    \end{subcaption}
    
    \begin{subcaption}{Validation set}
        \begin{tabular}{c|c c c}
         \hline
         \makecell{Slide \\ HER2 score} & Precision & Recall & Dice score\\
         \hline
         $0$ &  $.903 \pm .016$ & $.897 \pm .053$ & $.900 \pm .034$\\
         $1+$ &  $.864 \pm .093$ & $.916 \pm .030$ & $.888 \pm .064$\\
         $2+$ &  $.906 \pm .013$ & $.960 \pm .013$ & $.932 \pm .002$\\
         $3+$ &  $.897 \pm .024$ & $.921 \pm .024$ & $.909 \pm .003$\\
         \hline
    \end{tabular}
    \end{subcaption}
    
    \begin{subcaption}{Testing set}
        \begin{tabular}{c|c c c}
         \hline
         \makecell{Slide \\ HER2 score} & Precision & Recall & Dice score \\
         \hline
         $0$ &  $.822 \pm .105$ & $.839 \pm .107$ & $.822 \pm .025$\\
         $1+$ &  $.841 \pm .081$ & $.905 \pm .014$ & $.870 \pm .037$\\
         $2+$ &  $.909 \pm .012$ & $.937 \pm .013$ & $.923 \pm .003$\\
         $3+$ &  $.845 \pm .024$ & $.938 \pm .025$ & $.888 \pm .006$\\
         \hline
    \end{tabular}
    \end{subcaption}
    \label{tab:segmentation_results}
\end{table}

\subsection{HER2 Patch classification}
\begin{figure}[h]
    \centering
    \includegraphics[width=0.8\linewidth]{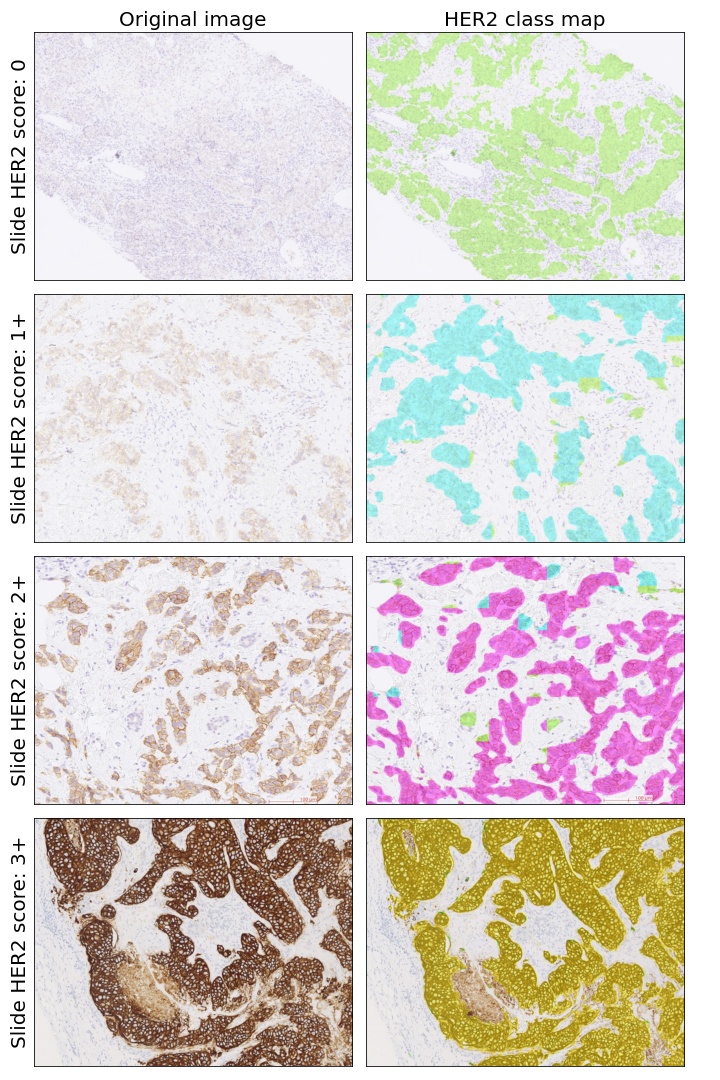}
    \caption{Visualization of the HER2 patch classification model output. Class 0 patches are represented in green, 1+ patches in blue, 2+ in pink and 3+ in yellow.}
    \label{fig:calopix}
\end{figure}

\begin{figure}[h]
    \centering
    \includegraphics[width= 0.9\linewidth]{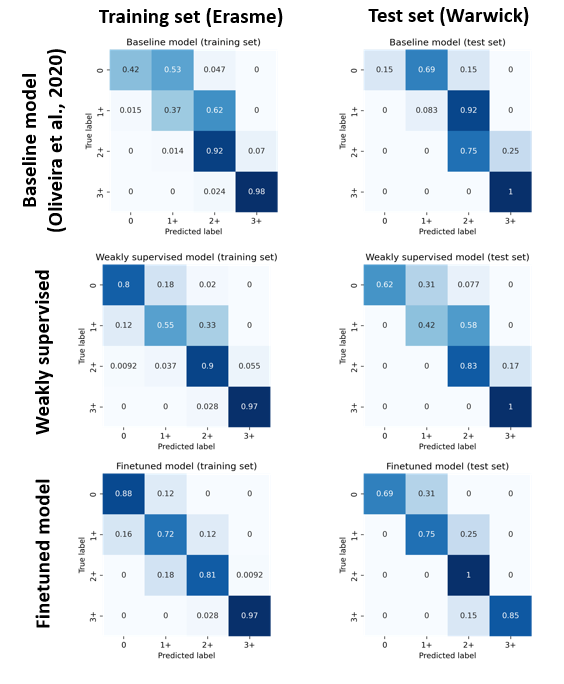}
    \caption{Confusion matrices for each step of our end-to-end framework for scoring HER2 slides. The confusion matrices show the performances on the training set (left column) and test set (right column). Every stage improves the metrics over the previous one.}
    \label{fig:classifMetrics}
\end{figure}

\begin{figure}[h]
    \centering
    \includegraphics[width= 0.9\linewidth]{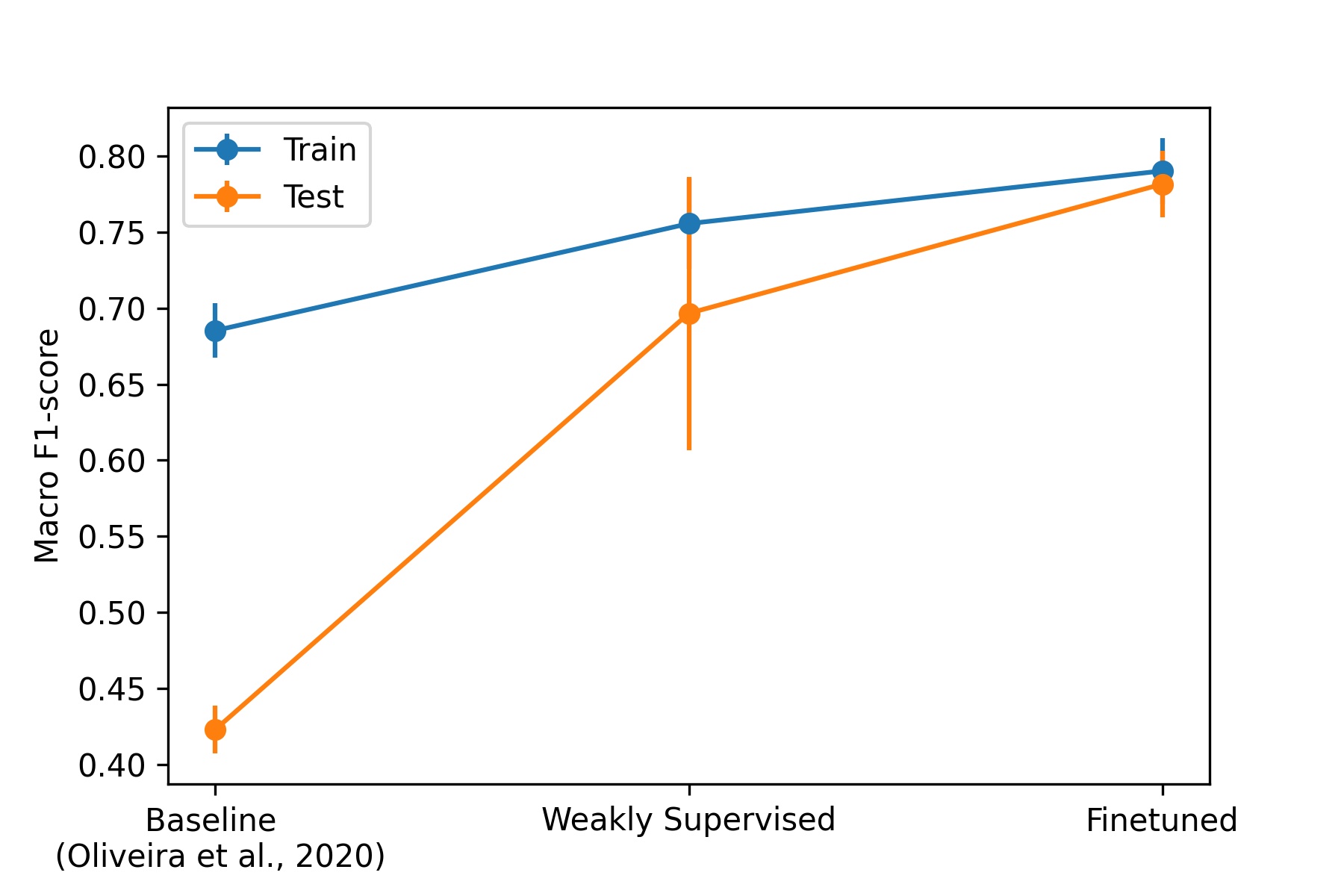}
    \caption{Macro-averaged F1-score for each step of our end-to-end framework for scoring HER2 slides. Every stage improves the metrics over the previous one, the final stage yielding a macro F1-score of 0.78 on the hold-out test set.}
    \label{fig:classifMetrics2}
\end{figure}

To show the effect of our 3-step approach, we evaluate our pipeline at each stage. For each stage, we compute the macro F1-score and the confusion matrix of the model on the training and test sets (see Figures \ref{fig:classifMetrics} and \ref{fig:classifMetrics2}). The baseline model trained in the same way as \citet{oliveira2020weakly}, which is used as pretraining, already performs well on HER $2+$ and $3+$ classes, with F1-scores above $0.92$ both on the training and testing set. However, it overclassifies slides with lower HER2 scores, 0 slides being classified as 1+, and 1+ as 2+. This issue is corrected by the weakly supervised training, especially for class 0 slides, but there is still some confusion for 1+ slides. The use of the clinical guidelines as linear constraints in the weakly supervised training prevents the model from overclassifying patches as the amount of patches of each class are constrained by the slide class: a slide cannot have too many patches from higher classes, which results in better predictions at the slide level. The final supervised finetuning on the logits improves the model's prediction for lower classes, especially 1+ slides. Figure \ref{fig:calopix} shows some inference results on slides of different HER2 score. \\


\subsection{Using our end-to-end pipeline to study HER2 class hetereogenity}

\begin{figure}[h]
    \centering
    \includegraphics[width=\linewidth, trim={4cm 2cm 4.5cm 2cm},clip]{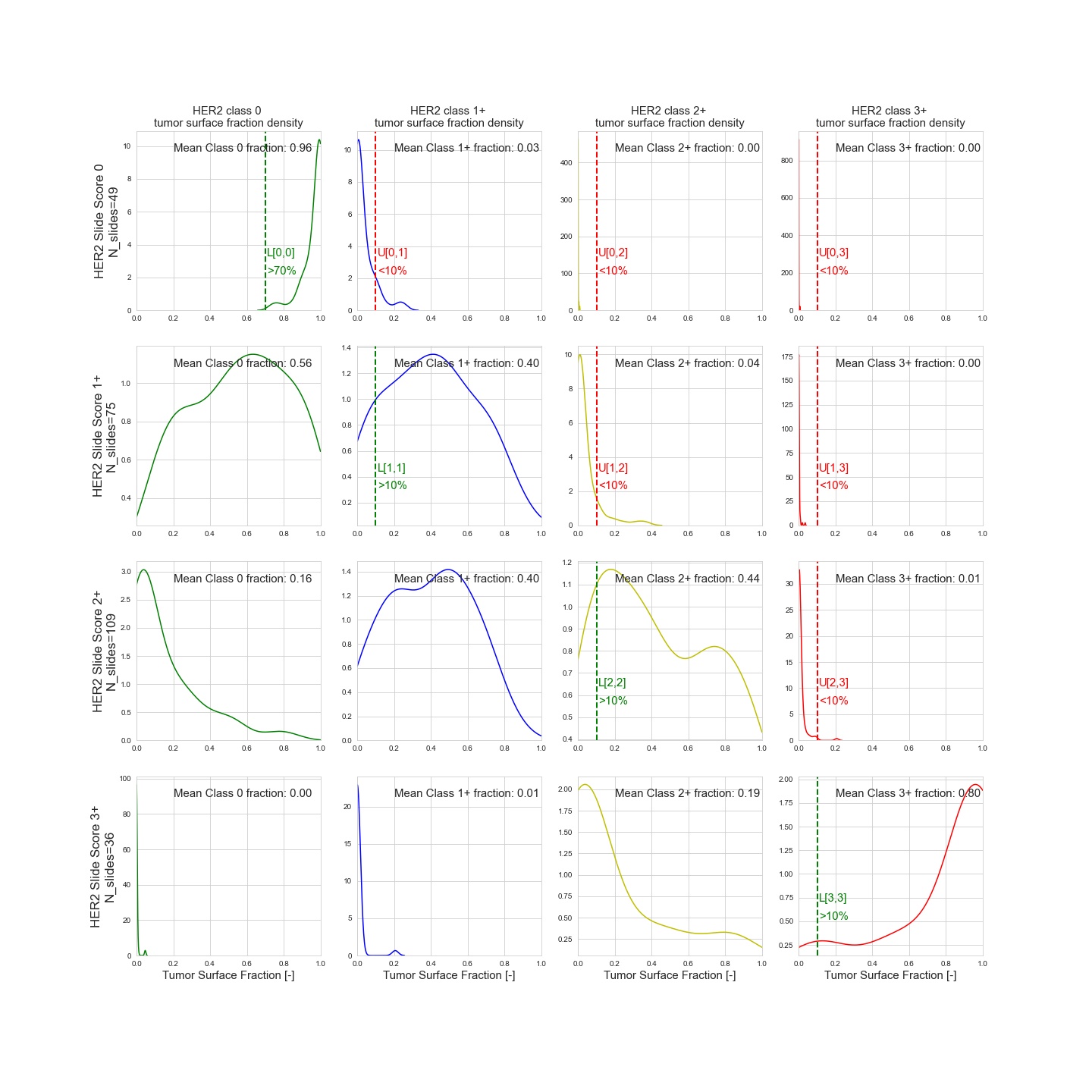}
    \caption{Kernel Density Estimations (KDE) for each HER2 class tumor surface fraction by slides' HER2 scores on the training set. Each row represents a slide HER2 score and each column the patches HER2 scores. For slides of a known HER2 score, the tumor surface fraction of lower HER2 classes (corresponding to graphs below the diagonal) do not follow particular distributions as there are no constraints on these fractions. For higher HER2 classes (corresponding to graphs above the diagonal), our weakly constrained optimization enforces their surface fraction to be below $10 \%$}
    \label{fig:kde}
\end{figure}

\begin{figure}[h]
    \centering
    \includegraphics[width=\linewidth, trim={4cm 2cm 4.5cm 2cm},clip]{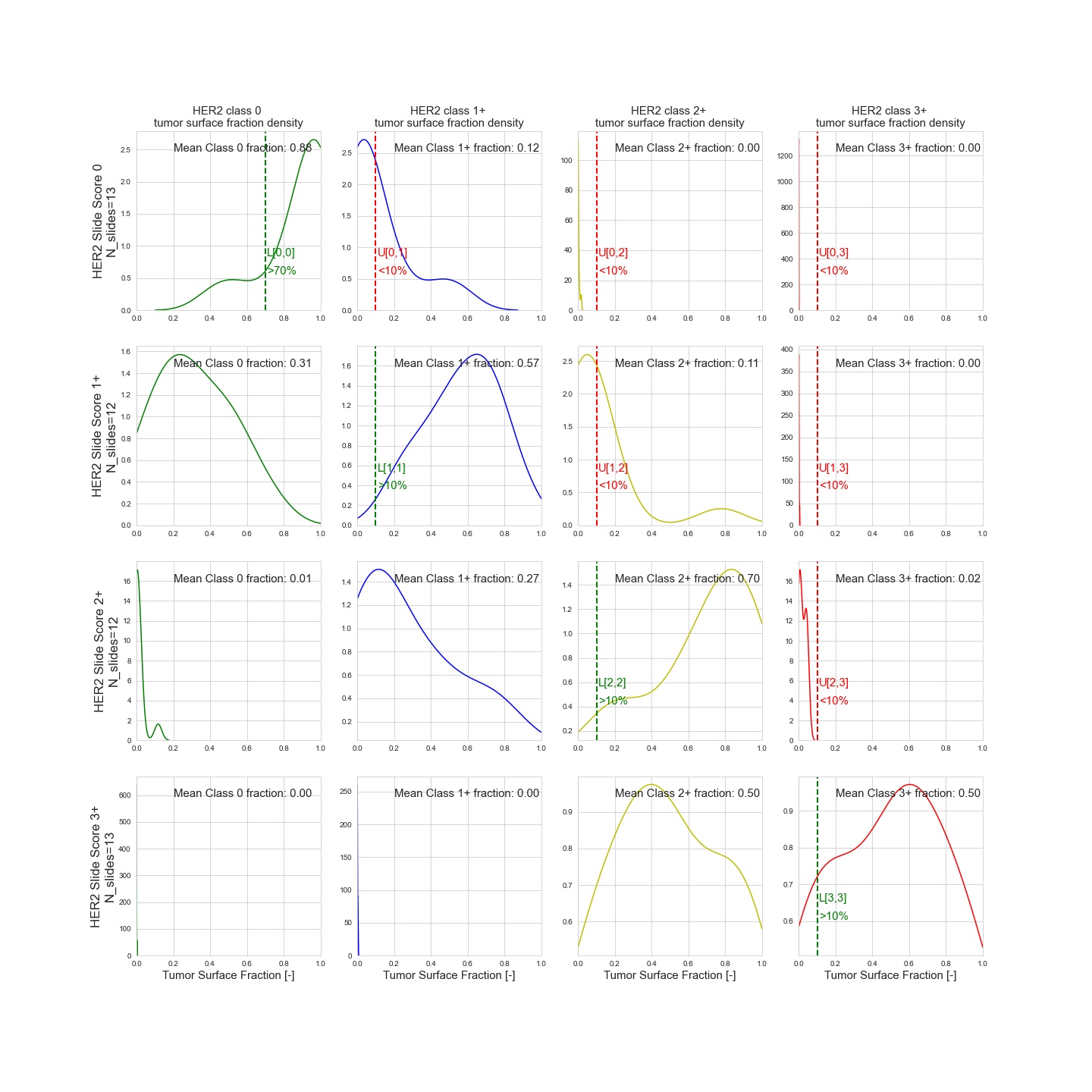}
    \caption{Kernel Density Estimations (KDE) for each HER2 class tumor surface fraction by slides' HER2 scores on the Warwick training set, which we use as a hold-out test set.}
    \label{fig:kde2}
\end{figure}

In order to interpret the predictions of our model, we are interested in the distributions of the different HER2 phenotypes predicted within every slides of in the training and testing datasets. The purpose is to see how much of tumor surface of different HER2 patch classes is present in slides that were classified as a certain overall HER2 slide score. Thus, we plot a Kernel Density Estimation (KDE) of the tumor surface fraction for each phenotype grouped by slides' HER2 scores. The result are shown in Figure \ref{fig:kde} for the training set and Figure \ref{fig:kde2} for the test set. Each row represents the slides' HER2 score and each column the patches grouped by HER2 score. The dotted lines are the constraints dictated by the ASCO/CAP clinical guidelines (described in Table \ref{guidelines} and Equation \ref{constraint_matrices}). The graphs below or above the diagonal represent the proportion of invasive carcinoma from a lower or higher class than the slide's known class respectively. \\
In the first row, which corresponds to slides with a HER2 score of 0, one can see that most slides have more than 70 \% of their invasive carcinoma classified as 0. Although there are still a few slides with some non-negligible fraction ($ \geq 10 \%$) of $1+$ invasive carcinoma, the fraction of higher classes invasive carcinoma is concentrated below $10 \%$, which corresponds to the threshold defined by the ASCO/CAP clinical guidelines. \\
In general terms, for slides with a score between 0 and $2+$, the graphs above the diagonal show the effect of the upper linear constraints derived from the clinical guidelines, which limit the amount of invasive carcinoma classified as higher HER2 classes: all of the invasive carcinoma surface fractions are concentrated below $10 \%$. \\
For 1+ and 2+ slides, the lower-classes tumor surface fraction distributions are spread out, which highlights the heterogeneity within these classes. Heterogeneous slides represent hard cases for pathologists as the risk of error due to the selection of the region of interest is more significant. On the contrary, $3+$ slides are very homogeneous as there are almost no 0 and $1+$ invasive carcinoma in these slides. \\

\section{Discussion}
\label{discussion}
In this work, we proposed and implemented a constrained weakly supervised approach for HER2 scoring. For the sake of interpretability, we chose to directly implement in our pipeline the ASCO/CAP guidelines based on the tumor surface percentages. We use the surface as a proxy for the number of cells, which we argue is reasonable since cell size usually does not vary a lot within the same slide. Our pipeline contrasts with other approaches for HER2 scoring like \citet{chen2021diagnose} who use an aggregation model on top of their weakly supervised patch classification model. Thus, they no longer follow the clinical guidelines but directly optimize their network on the slides' labels which are noisy as shown by \citet{Thomson2001},  \citet{Hoang2000} and Figure \ref{fig:pat1}. Indeed, we found that there was a $30\%$ discordance rate between pathologists 1 and 2. \\
As we conducted a multi-pathologist consensus to mitigate the label noise of the slides used for training the model, there were some cases where both pathologists were hesitant and used the FISH results to guide their decision, which is a piece of information that the model does not use. For these hard cases, the interpretability of our model acts as a second opinion, by providing useful insight into the proportions of invasive cancer surface for each HER2 score.  \\

As the first step in our pipeline, the invasive carcinoma segmentation model achieves an average Dice score of above 0.91 on the test set on slides of class $2+$ and $3+$. The results for classes 0 and $1+$ were slightly worse, at Dice scores $0.82$ and $0.87$ respectively. These slides were also the hardest slides to annotate, which could also impact negatively the results as stated by \citet{vuadineanu2021analysis}. \\

Both invasive carcinoma segmentation and the HER2 patch classification models process images at a magnification of $10 \times$ ($1 \mu m / pixel$) contrary to other studies that work at $20 \times$ or even $40 \times$ such as \citet{vandenberghe2017relevance} or \citet{chen2021diagnose}. To compare our results, we group $0$ and $1+$ slides in the same category (HER2-negative) as they do. They achieve a macro-averaged F1-score of 0.751 on the whole data cohort and 0.907 on four-fold cross-validation respectively. Despite processing the image at half or a quarter of their resolution respectively, we reach similar or better performances as we achieve a macro-averaged F1-score of 0.887 on our hold-out test set. Working at a lower magnification induces a faster inference time, so that using our solution does not reduce the pathologist's working speed.  \\ 
As for our proposed workflow, shown in Figure \ref{fig:graphical_abstract}, we still require a pathologist to draw a ROI on the slide. One improvement would be to make the segmentation model able to segment invasive carcinoma from carcinoma in situ and benign stained tissue in addition to benign tissue. On Hematoxylin and Eosin (HE) stained slides, \citet{kanavati2022deep} built a model to segment invasive carcinoma in situ by using a Convolutional Neural Network (CNN) followed by a Recurrent Neural Network (RNN). Adding this step would make our pipeline fully automatic and more accurate as the whole invasive cancer surface will be taken into account to determine the slide's HER2 score. \\
To do HER2 scoring on segmented invasive carcinoma, we decide experimentally to keep patches with more than $10 \%$ of invasive carcinoma surface for the HER2 patch classification step. This allows us to treat $96 \%$ on average of the invasive cancer within the drawn ROI. In particular, patches with little invasive cancer surface are the ones on the boundary of cancer tissue or isolated cancer cell clumps, and thus contain a lot of benign tissues. Experimentally, we found that these patches tend to be under-classified compared to the HER2 score the cancerous tissue should have even though they were specifically included in the training. Although improvements can still be made for isolated infiltrating cancer patches, we observed that this issue does not significantly impact the global HER2 scoring. Considering HER2 scores 0 and $1+$ as two different classes, we finally obtain a macro-average F1-score of 0.78 in predicting the slides' HER2 scores on the hold-out test set and only make mistakes on adjacent classes (see Figures \ref{fig:classifMetrics} and \ref{fig:classifMetrics2}). On the training set, where the true slide labels were obtained through a multi-pathologist consensus, the model achieves a macro-average F1-score of 0.80, where pathologist 1 achieves an F1-score of 0.71 and pathologist 2 (who participated in the final consensus meeting) an F1-score of 0.91. \\

\begin{figure}[h]
    \centering
    \includegraphics[width=0.9\linewidth]{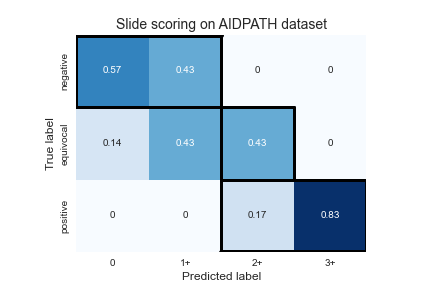}
    \caption{Confusion matrix on AIDPATH dataset used as a test set. Note that we only know the clinical outcome for this dataset (HER2-negative, HER2-equivocal, and HER2-positive). There are only misclassification for equivocal slides. The cells with thick borders represent correct predictions : HER2-negative corresponds to classes 0 and $1+$, HER2-equivocal to class $2+$, and HER2-positve to classes $2+$ and $3+$.}
    \label{fig:cm_aidpath}
\end{figure}

To verify the generalization of our model to different domains (see Figure \ref{fig:patches}), we evaluate our pipeline on the AIDPATH dataset as the other datasets were scanned with scanners all from Hamamatsu. The scanner for the AIDPATH dataset was unknown but the stain expression is visibly different from our training set as shown in Figure \ref{fig:patches}. As this dataset's labels were HER2-negative, equivocal and HER2-positive, we grouped the slides with a predicted class of 0 or $1+$ slides together in the HER2-negative class. We get a macro-averaged F1-score of 0.77. Figure \ref{fig:cm_aidpath} shows that the only error occurs for equivocal slides, whereas HER2-negative and positive slides are all well classified. 

\begin{figure}[h]
    \centering
    \includegraphics[width=0.8\linewidth]{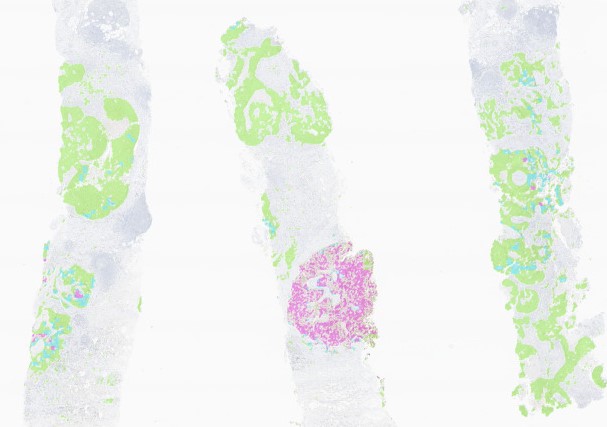}
    \caption{HER2 class map generated by our model. Invasive cancer of HER2 class 0 is represented in green, class $1+$ in blue, and class $2+$ in pink. The surface of $2+$ tumor is just below $10\%$, classifying the slide as 0 although it is closer to a $2+$ from a clinical point of view. The visualization allows the pathologist to understand quickly the decision of the model.}
    \label{fig:heterogeneous}
\end{figure}

\subsection{Analysis of rare heterogeneous HER2 Slides}\label{slideAnalysis}

In the clinical guidelines, rarely occurring heterogeneous HER2 slides are those which contain a nonzero - but less than 10\% - tissue fraction of an HER2 class which is two or more classes higher than the class it would be given if we were to directly evaluate the clinical guidelines in Table \ref{guidelines}. For instance, for the slide shown in Figure \ref{fig:heterogeneous}, the predicted HER2 class surface fractions are $[82\%, 9\%, 9\%, 0\%]$. The predicted class according to the principal guidelines is 0, whereas the clinical guidelines recommend that the slide be classified as $1+$ so that under-treatment of the patient is avoided. In the same way, heterogeneous slides that have a nonzero but less than 10\% fraction of $3+$ invasive carcinoma should be classified as $2+$, regardless of the proportion of the other classes. For such cases, we believe that the spatial class map interpretability that our model provides can be helpful for pathologists, especially for borderline cases. \\

\section{Conclusion}
In this paper, we presented an interpretable weakly supervised constrained deep learning model for HER2 scoring. We directly leveraged the ASCO/CAP guidelines, both as constraints for training our model, and for inference, to compute the slide's class from the classes of the patches. Throughout our work, we focused on the interpretability of our model, for the pathologist especially, by outputting a HER2 class map along surface percentages for the invasive cancer within the slide. By studying the distribution of the tumor surface percentages of each HER2 score, we were also able to quantify HER2 intra-class heterogeneity, leading to a better understanding of the inter-observer variability in HER2 scoring.  

\section*{Acknowledgements}
This study was approved by the Ethical Committee of the Erasme University Hospital (P2021/512). According to Belgian law, no written informed consent was required for archival material in the context of retrospective studies. This work was carried out with the support of the ``Fonds Yvonne Bo\"{e}l'' (Brussels, Belgium). We also thank Professor Marc Aubreville for his insightful discussions and comments revising this article.

\bibliographystyle{model2-names.bst}\biboptions{authoryear}
\bibliography{refs}

\end{document}